# Inter-layer spin diffusion and electric conductivity in the organic conductors κ-ET$_2$-Cl and κ-ET$_2$-Br

[1]Ágnes Antal, [1]Titusz Fehér, [1]Erzsébet Tátrai-Szekeres, [1]Ferenc Fülöp, [1,2]Bálint Náfrádi, [2]László Forró, [1]András Jánossy

[1]*Institute of Physics, Budapest University of Technology and Economics, and Condensed Matter Research Group of the Hungarian Academy of Sciences, P.O.Box 91, H-1521 Budapest, Hungary*
[2]*Institute of Condensed Matter Physics, Ecole Polytechnique Fédérale de Lausanne, CH-1015 Lausanne, Switzerland*

A high frequency (111.2–420 GHz) electron spin resonance study of the inter-layer (perpendicular) spin diffusion as a function of pressure and temperature is presented in the conducting phases of the layered organic compounds, κ-(BEDT-TTF)$_2$-Cu[N(CN)$_2$]$X$ (κ-ET$_2$-$X$), $X$=Cl or Br. The resolved ESR lines of adjacent layers at high temperatures and high frequencies allows for the determination of the inter-layer cross spin relaxation time, $T_x$ and the intrinsic spin relaxation time, $T_2$ of single layers. In the bad metal phase spin diffusion is two-dimensional, i.e. spins are not hopping to adjacent layers within $T_2$. $T_x$ is proportional to the perpendicular resistivity, $\rho_\perp$, at least approximately, as predicted in models where spin and charge excitations are tied together. In κ-ET$_2$-Cl, at zero pressure $T_x$ increases as the bad metal–insulator transition is approached. On the other hand, $T_x$ decreases as the normal metal and superconducting phases are approached with increasing pressure and/or decreasing temperature.

## 1. Introduction

Organic layered conductors, and in particular members of the κ-(BEDT-TTF)$_2$Cu[N(CN)$_2$]$X$ (henceforth κ-ET$_2$-$X$) family, are strongly interacting electronic systems on the borderline of a metal to insulator transition. The crystal structure [1],[2] (Fig.1) consists of conducting or magnetic ET molecular layers separated by insulating single-atom-thick Cu[N(CN)$_2$]$X$, $X$=Cl or Br polymer sheets. The electronic bands are usually considered half-filled (with 1 hole per molecular dimer, ET$_2$), narrow and highly anisotropic. The intermolecular overlap integrals of neighbouring molecules within a layer and in-between layers are of the order of $t_{//} \approx 0.1$ eV [3] and $t_\perp \approx 0.1$ meV [4], respectively. Electric conductivity is two-dimensional (2D), measured anisotropies are between 100 and 1000. At low temperatures, under moderate pressures, a Fermi liquid description is thought to be valid for the in-plane transport. However, the perpendicular conductivity is at most "weakly coherent", tunneling to the second neighbour layer is incoherent even at low temperatures. At higher temperatures the parallel conductivity is also incoherent: the electronic mean free path is much less than the molecular separations.

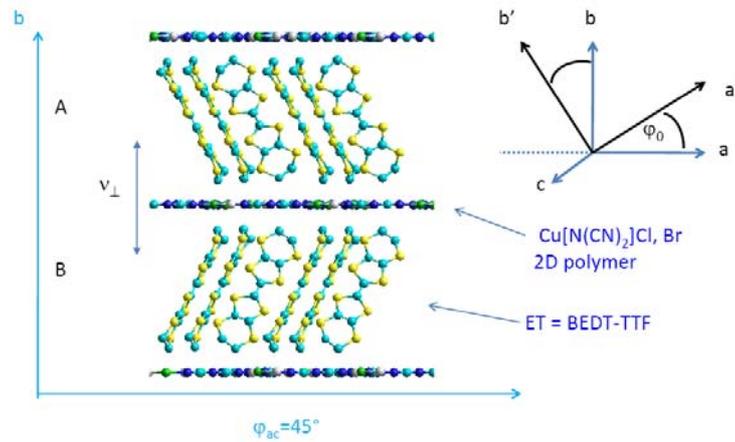

Fig.1. Structure of κ-ET$_2$-Cl and κ-ET$_2$-Br. Conducting ET (bis(ethylenedithio)-tetrathiafulvalene) layers are separated by insulating Cu[N(CN)$_2$]$X$, $X$=Cl or Br polymeric layers. $\nu_\perp$ is the inter-layer electron hopping rate measured by ESR. $\varphi_{ac}$ denotes the angle from $a$ in the ($a$, $c$) plane. The Larmor frequencies of chemically equivalent A and B layers are different in general orientation magnetic fields. $\varphi_0$ , $a$', $b$' indicate the orientation of the g-tensor principal axes of layer A.

The pressure-temperature phase diagram [5],[6],[7] is of particular interest. κ-ET$_2$-$X$ is a model system, in which the magnetically ordered and the various conducting phases are reached with relative ease (Fig.2). The high temperature phase is a "bad metal" characterized by a very high resistivity. κ-ET$_2$-Cl, at ambient pressures undergoes a Mott–Hubard metal–insulator transition below 50 K. It is a weak antiferromagnet below $T_N$=26 K. In κ-ET$_2$-Cl at moderate pressures and in the isostructural system, κ-ET$_2$-Br there is a "bad metal" to normal metal transition below about 50 K. In these, the ground state is superconducting at pressures up to about 0.5 GPa. At even higher pressures the ground state is a normal metal.

Here we measure the inter-layer spin diffusion rate in κ-ET$_2$-$X$, $X$=Cl, Br as a function of temperature, and in κ-ET$_2$-Cl, also as a function of pressure. The method relies on the analysis of the conduction electron spin resonance (CESR) spectra at high magnetic fields and yields high precision data for the hopping rate. This is in contrast with conductivity measurements where the quality of electric contacts and the inhomogeneity of the current distribution make absolute values uncertain. Our method makes use of the alternating sandwich structure of κ-ET$_2$-Cl and is not applicable to all layered systems. Nevertheless, it has a wide application range and it can be used in various systems with weakly interacting non-equivalent magnetic sublattices. It has been applied to 1D chain-like systems, many years ago [8]. With the availability of highly sensitive high frequency ESR spectrometers, there are many more interesting systems that can be studied.

We find that spin diffusion is 2D at high temperatures. In the metallic compounds a 2D to 3D crossover occurs at lower temperatures. Spin diffusion in quasi classical models is proportional to the electrical transport and the density of states of the metallic layers. Indeed, we find that this is the case, at least qualitatively: the temperature and pressure dependence of the interlayer spin hopping rate follows that of the perpendicular conductivity.

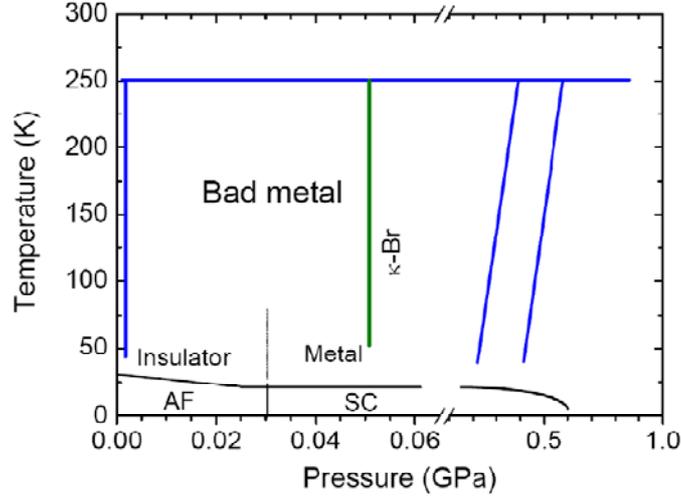

Fig.2. Schematic pressure–temperature phase diagram of κ-ET$_2$-Cl. ESR experiments on κ-ET$_2$-Cl were along thick lines. The κ-ET$_2$-Br ESR experiment line at ambient pressure is marked at 50 MPa to take into account the "chemical pressure".

The paper is organized as follows. Following a brief description of the experimental setup and samples (Sec. 2) we detail in Sec. 3 the method used to determine the inter-layer coupling parameters. Sec. 4 is devoted to the experimental results, we present the spin hopping times between layers, T$_x$, and the intrinsic spin life time of individual layers, T$_2$. In Sec. 5 we compare the spin hopping times and conductivity. A short account of the method was published in Ref. [9].

## 2. Experimental methods

κ-ET$_2$-Cl and κ-ET$_2$-Br single crystals with typical dimensions of 1×1×0.2 mm$^3$ were grown by standard electrochemical methods in Ar gas filled electrolytic cells. The electrolytic cell was placed in an additional temperature regulated argon gas chamber to prevent oxygen and water contamination. The quality of several single crystals was verified by X-ray diffraction.

The high frequency ESR spectrometers at EPFL (210, 315 and 420 GHz) and BUTE (111.2 and 222.4 GHz) have similar designs [10] [11] [12]. A non-resonant mm-wave circuit allows for *in situ* change of the frequency and rotation of the crystal around a single axis. An oil filled clamp pressure cell was used. The pressures quoted are nominal values at 300 K; the pressure loss between 300 and 4 K is about 0.2 GPa. The sample cooling rate was 1 K/min or slower. Some samples were warmed and cooled several times but there was no thermal or magnetic field dependence in the ESR spectrum.

Four probe dc conductivity measurements in the highly conducting planes were done with gold contacts attached to the thin crystal surfaces. The temperature sweeps deteriorated the contacts and the first down sweep data are reported.

## 3. Measurement method of inter-layer spin diffusion and exchange fields

Weak interactions between magnetic sublattices can be precisely measured from an analysis of the ESR spectrum [13]. The method is applicable whenever there are at least two weakly interacting magnetic sublattices with different Larmor frequencies. There are a large number of organic crystals with alternating layered or molecular chain structures. Very small interactions are measurable, e.g. in the antiferromagnetic phase of κ-ET$_2$-Cl an inter-layer effective magnetic field of 1 mT was found [14]. This value is several orders of magnitude smaller than exchange fields within a single layer. Here we apply the method to measure the magnetic interactions, the hopping rate and exchange fields, between adjacent ET molecular layers in the paramagnetic phases of κ-ET$_2$-X, X=Br and Cl. The orthorhombic crystal structure of κ-ET$_2$-X ([1],[2]) maybe looked upon as a sandwich of two monoclinic crystals (Fig.1) in which the orthogonal $a$ and $c$ axes coincide while the $b_A$ and $b_B$ axes of the A and B layers are tilted from the orthorhombic $b$ axis. The conducting ET layers are separated by a single-atom-thick Cu[N(CN)$_2$]X "insulating" polymeric layer. The interaction between the layers is determined from the mixing of the ESR spectra of the separate layers. The interlayer hopping frequencies are between $10^{-10}$ s and $10^{-8}$ s, this interaction is weak, but essential for the understanding of the electronic and magnetic properties.

*3.1 g-factor anisotropy*

In a general direction magnetic field, **B**, the Larmor frequencies, $\nu_A$ and $\nu_B$ of the chemically equivalent but crystallographically different adjacent ET layers, A and B are different. There are to a first approximation two Lorentzian ESR lines in a magnetic field tilted by $\varphi_{ab}$ from $a$ in the ($a$, $b$) plane if interactions between layers are weak:

$$\nu_A = (\mu_B/h)\, g_A(\varphi_{ab})\, B = \gamma_A/(2\pi)\, B \qquad (3.1)$$

$$\nu_B = (\mu_B/h)\, g_B(\varphi_{ab})\, B = \gamma_B/(2\pi)\, B$$

As usual, we denote by $g_A(\varphi_{ab})B$ the absolute value of the vector $g_A\mathbf{B}$. The principal values of the $g$ tensor are $g_{Ai}$, $i=a, b, c$. The principal directions of the $g$-factor tensors are rotated about the $c$ axis by $\pm\varphi_0$ from the orthorhombic $a$ axis. $g_A$ and $g_B$ depend mostly on the ET molecular properties [15] and depend little on temperature (Fig. 3). (In κ-ET$_2$-Cl below 50 K antiferromagnetic fluctuations decrease the resonance frequency). The ESR is split in the ($a$, $b$) plane, while $\nu_A = \nu_B$ in the ($a$, $c$) and ($c$, $b$) planes in the paramagnetic phase. (In the antiferromagnetic phase, interactions between the planes split the resonance in these planes also [14]).The $g$-factor anisotropies of the A and B layers in κ-ET$_2$-Cl and κ-ET$_2$-Br at high temperature (250 K) and high magnetic fields, ($\nu_L$= 222.4 GHz, B ≈ 8 T) have the usual sinusoidal form. The A and B layer lines follow each other with an angle difference of $2\varphi_0$ (Figure 2). The principal values are: $g^m_{a'}$= 2.0048 (2.0048), $g^m_{b'}$ =2.0099 (2.0102), $\varphi_0 = 32^0$ ±5$^0$ (29$^0$ ±5$^0$), $g^m_{c'} = g_c$ = 2.0050 (2.0049) in the Cl (Br) crystals ($g^m_{i'}$ denote the $g$ factors of monoclinic layers). We estimate an error of ±0.0002 in the $g$ factors from a ±5$^0$ alignment uncertainty and the uncertainty in $g_{KC60}$ =2.0006 ±0.0001 of the polymeric KC$_{60}$ reference. Under the conditions of Fig. 2, the interactions between the layers affect little (although measurably) the ESR spectrum and the g-factors were obtained by fitting two equal intensity Lorentzians with different $g$ factors and line widths, $1/(\gamma_A T_{2A})$ and $1/(\gamma_B T_{2B})$.

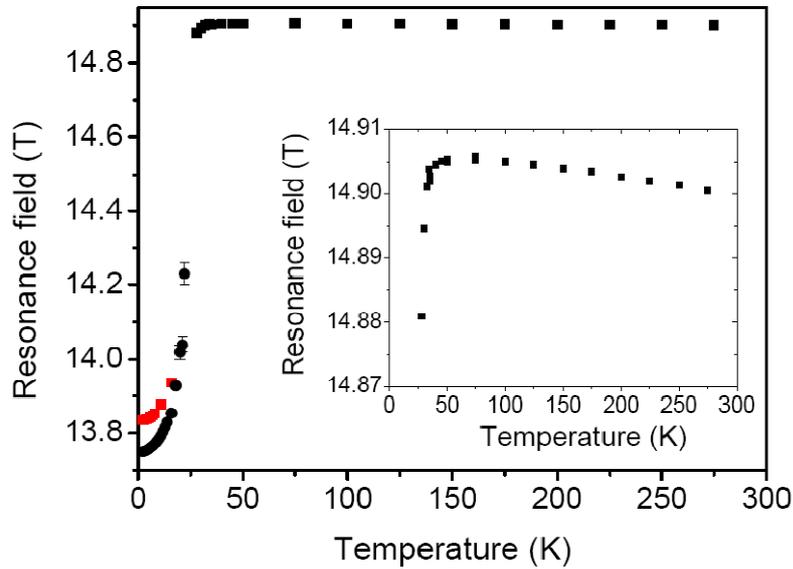

Fig.3. ESR field, $B_0$ versus temperature in κ-ET$_2$-Cl at 420 GHz exciting frequency. **B$_0$** is along the *b* crystallographic direction. Inset shows the small temperature dependence in the paramagnetic region on a magnified scale. The rapid decrease of $B_0$ below 50 K arises from 2D and 3D antiferromagnetic fluctuations. In the ordered phase, below $T_N$=26 K inter-layer interactions split the AFMR into two modes.

The resonance field is linear with frequency to high precision. At low Larmor frequencies there is a single „motionally narrowed" resonance line, the principal directions of the average g-factor tensor are along the orthogonal axes. The 9 GHz ESR *g* factors reported by Nakamura *et al*. [15] for κ-ET$_2$-Br: $g_a$=2.0063, $g_b$=2.0088, $g_c$=2.0048, are in agreement with our high frequency (222.4 GHz) values in the orthorhombic directions in κ-ET$_2$-Cl (κ-ET$_2$-Br): $g_a(\varphi_{ab}= 0) = 2.0061$ (2.0064), $g_b(\varphi_{ab}= \pi/2) = 2.0086$ (2.0088), $g_c$ =2.0050 (2.0049). The linearity of the Larmor frequency with field in κ-ET$_2$-Br is in contrast with the nonlinearity in the paramagnetic phase of the organic conductor EDT-TTF-CONMe$_2$AsF$_6$ attributed to the Dzyaloshinskii–Moriya interaction [16].

*3.2 The ESR spectrum of weakly interacting layers*

The interlayer spin diffusion and effective magnetic fields are measurable from the ESR spectra at sufficiently high magnetic fields whenever they are comparable to or stronger than the intrinsic spin relaxation rate of the layers. To calculate the ESR spectrum of the interacting layers, we assume that spin diffusion from the adjacent B layers increases the layer A magnetisation, ***M*** $_A$, with a rate of $(M_B – M_A)/T_x$. ***M***$_B$ also exerts a torque on ***M***$_A$ that can be expressed as an effective magnetic field ***B***$_{eff,AB}$ = $\lambda$***M***$_B$. Similar expressions with the same $T_x$ and $\lambda$ describe the effect of A layers on B.

The ESR spectrum (i.e. the dynamic susceptibility) is given by two coupled Bloch equations where the torque of the external static and exciting magnetic fields and the intrinsic spin relaxations of the layers are added. For layer A:

$$\frac{dM_{xy}^A}{dt} = \gamma^A (M^A \times M^B)_{x,y} - \frac{M_{xy}^A}{T_{2A}} + \frac{M_{xy}^B - M_{xy}^A}{T_x} \quad (3.2)$$

The coupled Bloch equations were solved numerically to obtain the absorption spectrum, $G(B)$ as a function of magnetic field magnitude for fixed Larmor frequency. The single layer parameters $g_A$, $g_B$, $T_{2A}$, $T_{2B}$ (or the average $T_2$) and the interaction parameters $\lambda M_0$ and $T_x$ are obtained from a best fit to the experiment ($M_0$ is the magnitude of the static magnetisation). Although in principle all parameters can be determined from the spectrum at a single Larmor frequency and sweeping the external field, $B_0$, the experiments were in most cases done at two or more frequencies without changing $\varphi_{ab}$. This gives a consistency check for the assumptions.

The spectrum consists of two Lorentzian lines if the line separation is much larger than the inter-layer interaction:

$$|\nu_A - \nu_B| \gg 1/T_x \text{ and } \gamma_0 \lambda M_0 \quad (3.3)$$

and there is a single line if

$$|\nu_A - \nu_B| \ll 1/T_x \text{ or } \gamma_0 \lambda M_0 \quad (3.4)$$

where $\gamma_0$ is the average gyromagnetic ratio. The interaction parameters are best determined when the splitting is of the order of $1/T_x$. The exchange field and cross relaxation have very different effects on the ESR spectrum. In conductors the cross relaxation dominates over the exchange fields and the critical value for two resolved lines to merge into a single line is

$$|\nu_A - \nu_B| \approx 1/T_x \quad (3.5)$$

The two lines merge smoothly as $1/T_x$ is increased from zero. The spectrum is symmetric around the frequency corresponding to the average $g$ factor at all values of $1/T_x$ if the intrinsic relaxation rates, $1/T_2$, are the same for both layers.

On the other hand, if $\lambda \neq 0$, i.e. for a magnetic coupling between the layers, the absorption spectrum $G(B)$ is asymmetric. For an antiferromagnetic coupling ($\lambda < 0$), the higher field resonance loses intensity while the lower field resonance gains intensity as $B_{eff,AB}$ increases from zero to a large value. Cabanas and Schwerdtfeger [8] realized this curious behaviour of the ESR in an insulating quasi 1D organic paramagnet with weakly interacting inequivalent chains.

## 4. Experimental results

*4.1. Bad metal phase, pressure dependence at 250 K*

At high temperatures, roughly above 50 K, $\kappa$-ET$_2$-X is in a bad metal phase, where the parallel conductivity is low and the parallel mean free path is less than typical intermolecular distances. In this phase the electronic conductivity increases with pressure [6], [17]. To measure the inter-layer spin hopping rate, the ESR spectra of a $\kappa$-ET$_2$-Cl single crystal (TEKCl8) were recorded at 250 K as a function of pressure between 0 and 1.06 GPa at 210, 315 and 420 GHz. The crystal was oriented with magnetic field along $\varphi_{ab} \approx 45^0$ in the (a,b) plane where the difference between the Larmor frequencies of the A and B layers is about the largest (Fig.4). The narrow AFMR lines in the antiferromagnetic phase [14] are an indication of the high quality of the crystal. Some of the measured and calculated spectra are displayed in Figure 5. An impurity Lorentzian ESR line from the pressure cell was subtracted with pressure independent positions and widths. The pressure was changed at ambient temperature in 40 MPa steps. The temperature dependence of the ESR was also measured at various fixed pressures. The ESR spectrum was independent of thermal or pressure history, in spite of the large number of cycles. The signal to noise ratio depends on pressure due to variations of the standing waves within the cell.

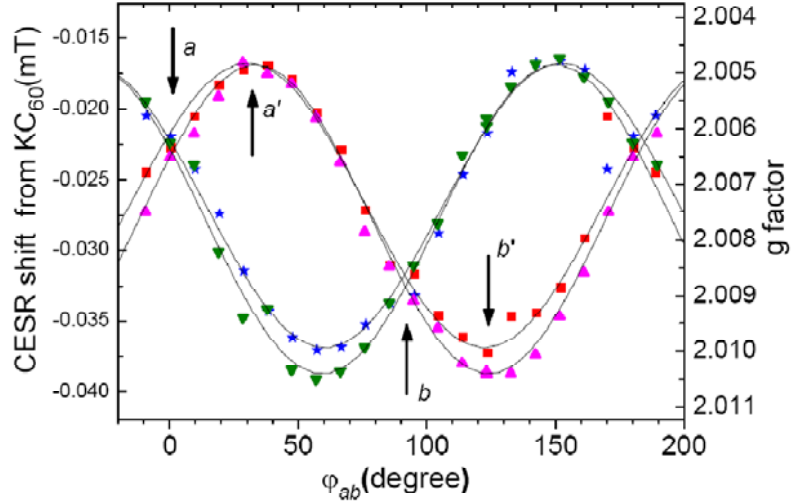

Fig. 4. *g*-factor anisotropies in the (a,b) plane measured at 222.4 GHz and 250 K. ET2Cl: red squares, ■ and blue stars, * and $\kappa$-ET$_2$-Br: magenta triangles up,▲; green triangles,▼. *a'* and *b'* denote g-factor principal axes of A-layers (see Fig.1), *a* and *b* are the orthorhombic lattice directions.

Fig. 5 shows the merger of A and B layer lines as the interlayer hopping rate, $1/T_x$ increases with pressure. We find $|\nu_A - \nu_B| \approx 1/T_x$ at 0.16 GPa and 0.32 GPa for 210 GHz and 420 GHz, respectively. The determination of $T_x$ is the most precise in this pressure range.

The pressure dependence of $T_x$ and the intrinsic relaxation time, $T_2$ derived from the ESR spectra are shown in Fig. 6. The calculated spectra are almost indistinguishable from the measured κ-ET$_2$-Cl spectra. The g factors were assumed to be pressure independent. We assumed layer independent relaxation rates, $T_{2A} = T_{2B} = T_2$, although a small difference between $T_{2A}$ and $T_{2B}$ would be probably more realistic. $T_x$, the cross relaxation time decreases by a factor of 20 between ambient and 0.94 GPa pressures. The intrinsic relaxation time, $T_2$ decreases smoothly by a factor of 2 in the same interval.

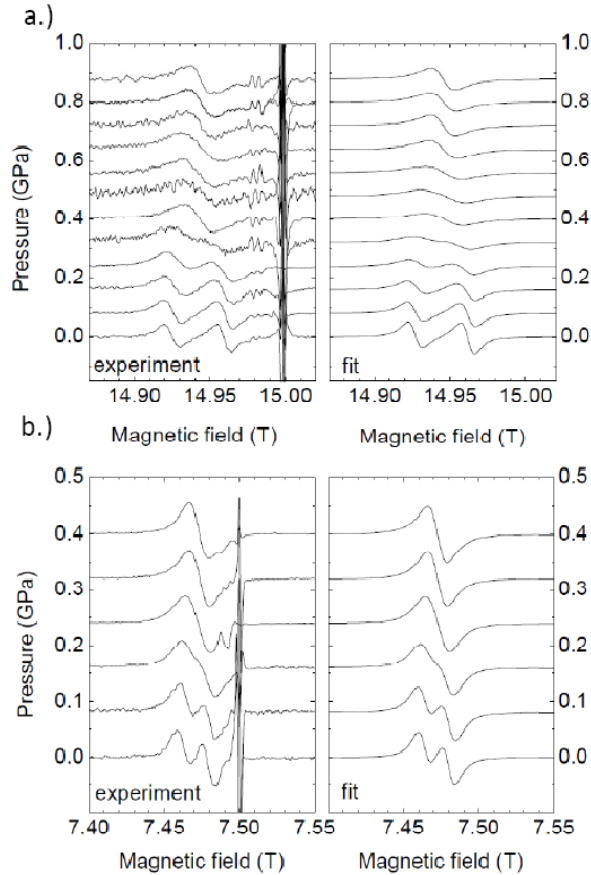

Fig.5. ESR spectra of κ-ET$_2$-Cl as function of pressure at 250 K. a.) $\nu_L$= 420 GHz, b.) $\nu_L$= 210 GHz. B oriented along $\varphi_{ab}$≈45° Left panels: experimental spectra. Right panels: best fit calculated spectra. Note the merger of A and B layer ESR lines with increasing pressure. The strong line at 15.00 T (7.50 T) is the KC$_{60}$ reference; an instrumental impurity line was subtracted at 14.98 T (7.49 T) at 420 GHz (210 GHz).

*4.2. Bad metal to normal metal crossover*

κ-ET$_2$-Br at ambient pressure and κ-ET$_2$-Cl under pressure undergo a continuous transition with decreasing temperature from a bad metal phase at ambient temperatures to a normal metal phase at low temperatures. κ-ET$_2$-Br has superconducting ground state and is regarded as a material similar to κ-ET$_2$-Cl but under a weak „chemical pressure" equivalent to about 50 MPa. κ-ET$_2$-Cl has a superconducting ground state at pressures between 30 to 500 MPa.

The crossover was observed in the ESR spectra as a function of temperature in three ET-Br crystals at ambient pressure and an κ-ET$_2$-Cl crystal at nominally p=0.4 and 0.64 GPa pressures. In the κ-ET$_2$-Br crystals the ESR changes little between 250 and 75 K. Below 50 K the A and B layer lines merge rapidly as the inter-layer hopping rate increases (Fig. 7.). There is a qualitative difference between the pressure dependence at 250 K in κ-ET$_2$-Cl and the temperature dependence at p=0 in κ-ET$_2$-Br (Figs. 6 and 7). The crossover as a function of temperature in κ-ET$_2$-Br involves an effective antiferromagnetic field **B**$_{eff,AB}$ between molecular layers. The A and B layer lines do not simply merge, the high field line loses intensity as the temperature is decreased below 50 K. Only a single line is left at 40 K.

The cross relaxation time, $T_x$, the intrinsic relaxation rates $T_{2A}$, $T_{2B}$ and the effective exchange field $\lambda M_0$ in κ-ET$_2$-Br between 40 and 250 K are shown in Figure 8. In the fits, $g_A$ was temperature dependent while the difference $g_A$–$g_B$ was fixed for all temperatures. The exchange field between layers is proportional to the external field. It is surprisingly large, $B_{eff0}$ = $\lambda M_0$ = 4 mT at 222.4 GHz and 40 K is larger than the (magnetic field independent) coupling between layers in the antiferromagnetic state of κ-ET$_2$-Cl.

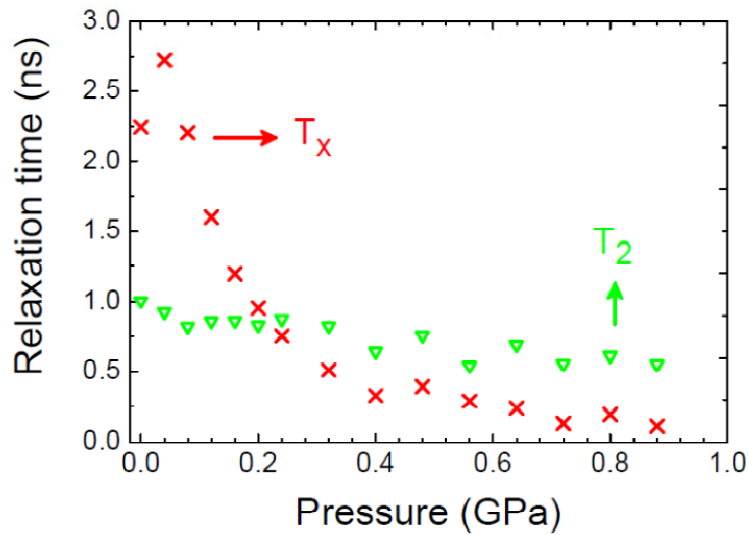

Fig.6. Interlayer cross relaxation time, $T_x$ and intrinsic relaxation time, $T_2$ as a function of pressure at 250 K in κ-ET$_2$-Cl determined from the fit to the ESR spectra in Fig.5. The crossover between 2D and 3D spin diffusion is at 0.2 GPa.

The increase of interlayer spin diffusion with decreasing temperature is well demonstrated in the κ-ET$_2$-Cl sample under pressures of 0.4 and 0.64 GPa. In these cases the hopping rate is fast and the g-factor anisotropy is insufficient to resolve the lines of layers A and B. $T_x$ in Fig. 9 was determined from the 420 GHz spectra, here $T_2$, was estimated from the broadening of the spectrum with increasing frequency at 250 K and was assumed temperature independent.

*4.3. Bad metal to insulator transition*

κ-ET$_2$-Cl has a continuous metal to insulator transition as the temperature decreases from 300 K to the Néel temperature at 26 K. The lengthening of the interlayer hopping time between 250 K and 50 K shows clearly the transition to the insulating state (Fig.10). The intrinsic spin relaxation, T$_2$ is approximately temperature independent above 50 K. Discussion of the complex behaviour between 26 K and 50 K (not shown), attributed to two and three dimensional fluctuations, is deferred to a later publication.

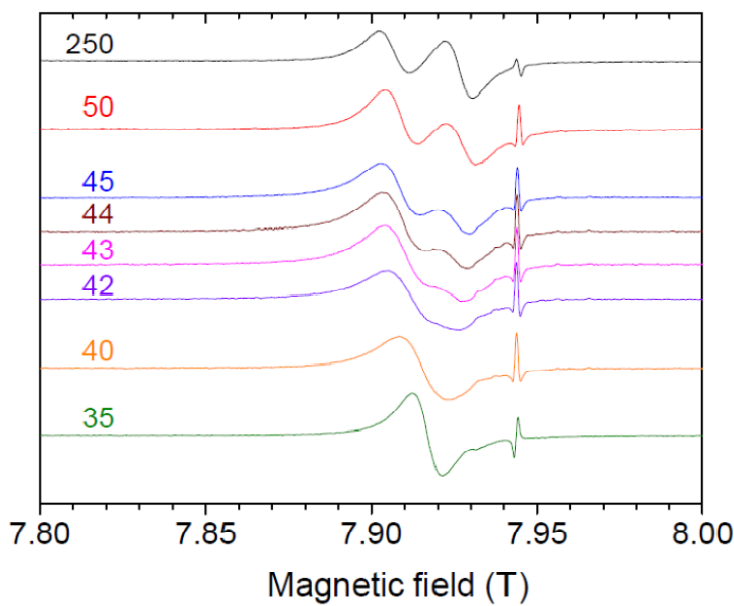

Fig.7. κ-ET$_2$-Br ESR spectra at 225 GHz as a function of temperature. B oriented along $\varphi_{ab} \approx 45°$. The spectra change little above 50 K. The A and B layer lines merge between 45 and 40 K.

*4.4. Inter-layer leakage*

A contribution to the perpendicular conductivity from leakage through defect holes several tens of nanometers apart in the polymeric layers has been proposed by McGuire *et al.* [18] for κ-ET$_2$-Br in an attempt to explain the temperature and frequency dependence of the conductivity anisotropy. This sample dependent effect was not taken into account in the fits discussed above. The excellent fit of the CESR spectra with well defined cross relaxation rates shows that such leaking does not play an important role. Leakage induces an inhomogeneous current distribution and locally increases spin diffusion at the defects while regions far from defects remain intact. The corresponding inhomogeneous broadening of the spectra has not been observed in the samples discussed here. Inhomogeneous line shapes, possibly due to leakage were observed, however in κ-ET$_2$-Br crystals from other growths.

## 5. Spin diffusion and electrical conductivity

*5.1 Relation between inter-layer spin hopping and electrical conductivity*

The determination of the interlayer charge hopping frequency, $\nu_\perp$ from the spin cross relaxation rate $1/T_x$ is the most important experimental result. We define $\nu_\perp$ as the charge hopping frequency through a *single* polymeric barrier

$$1/T_x = 2\,\nu_\perp . \quad (1)$$

$T_x$ arises from spin diffusion to the two neighbouring molecular layers. Strictly speaking, equation (1) is not always valid; spin cross relaxation can arise without charge transfer and charge transport is not necessarily accompanied by spin diffusion. Near the superconducting transition temperature fluctuating electron pairs tunnelling between layers contribute to charge transport but not to the spin transport. Magnetic and superconducting correlations are unimportant at temperatures well above the antiferromagnetic Néel temperature or the superconducting critical temperature and the quasi classical electron diffusion model is applicable, i.e. inter-layer spin and charge diffusion are tied together.

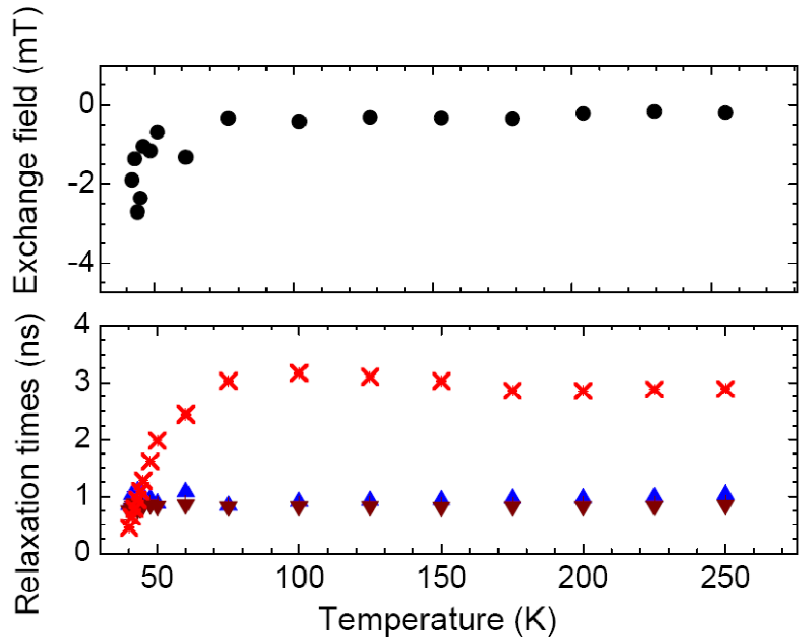

Fig.8. Relaxation times and exchange field in κ-ET$_2$-Br *versus* temperature calculated from the 222.4 and 111.2 GHz ESR spectra. Lower panel: interlayer cross relaxation $T_x$, (red crosses), intrinsic spin relaxations, $T_{2A}$ (triangles up) and $T_{2B}$ (triangles down). Upper panel: interlayer exchange field $B_{\mathrm{eff},0} = \lambda M_0$ at 8 T field. (see text). The shortening of $T_x$ below 75 K marks the onset of the bad metal to normal metal crossover. The 2D to 3D spin diffusion crossover is at 40 K.

In the metallic phase $\nu_\perp$ is closely related to the perpendicular electrical conductivity, $\sigma_\perp$. In a Fermi liquid at temperatures well below the Fermi energy, $E_F$ [19]

$$\sigma_\perp = e^2 \, g(E_F) \, d \, \nu_\perp / F, \qquad (2)$$

where $g(E_F)$ is the density of states for both spin directions of the metallic layers at $E_F$. $1/F$ is the two dimensional charge carrier density; in $\kappa$-ET$_2$-X, $F=(ac)/2$. Here $a$, and $c$ are the in-plane, $b=2d$ the out-of-plane lattice constants.

The hopping frequency, $\nu_\perp$ in Eq. (2), is from tunnelling between molecular layers and phonon assisted hopping over the barrier of the insulating polymeric layer. We assume that tunnelling through the barrier is the dominant mechanism for the perpendicular transport in the full temperature and pressure range of the experiments. As the temperature is raised, phonon assisted hopping over the barrier increases rapidly. In $\kappa$-ET$_2$-X the phonon contribution is certainly overshadowed by tunnelling at most temperatures and pressures, where $\nu_\perp$ increases with decreasing $T$.

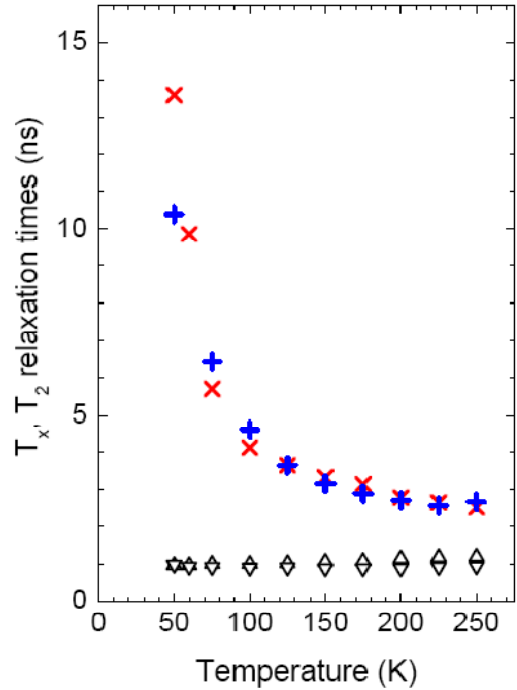

Fig.9. Cross spin relaxation, $T_x$ in two samples, (red x and blue +) and intrinsic spin relaxation times, $T_{2A}$ (up triangles) and $T_{2B}$ (down triangles) in $\kappa$-ET$_2$-Cl versus temperature above 50 K. The increase of $T_x$ follows the bad metal to insulator transition.

As we show, at ambient temperatures $\kappa$-ET$_2$-X is a "bad metal", the parallel mean free path within a single layer is much shorter than the inter-molecular distance and the concept of a density of states at a well defined Fermi energy fails (see below). The uncertainty of the momentum lifetime, $\hbar/\tau$ is comparable or larger than $E_F$ and in this case Eq. (2) is only approximately valid with $g(E_F)$ replaced by an average density of states. In a two dimensional free electron gas $g(E)$ is constant and this justifies that the average is about equal to $g(E_F)$.

*5.2 Comparison of perpendicular spin diffusion and electrical conductivity*

In this section we show that the expression (2) relating the perpendicular spin diffusion and perpendicular conductivity holds at least approximately all over the pressure-temperature phase diagram of κ-ET$_2$-X. We show that at ambient pressures, between 50 and 300 K, where the density of states changes little, $T_x$ and $\rho_\perp$ are proportional. Here the resistivity, $\rho_{\perp s}$ calculated from $T_x$ agrees with typical reported $\rho_\perp$ values. The agreement with $\rho_\perp$ data under pressure is only qualitative.

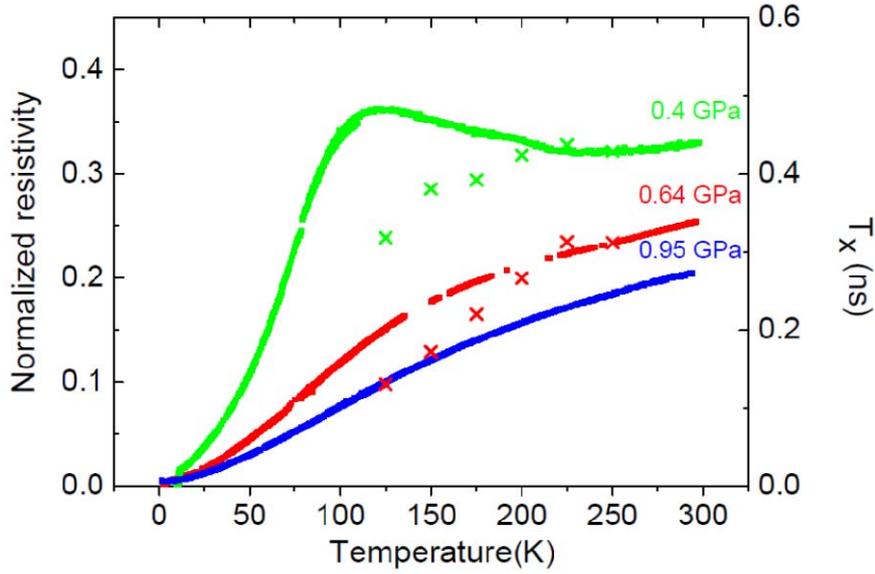

Fig. 10. Comparison of the cross relaxation, $T_x$ (crosses) and the parallel electrical resistivity (lines), in κ-ET$_2$-Cl under pressure. $\rho_{//}$ is normalized to the zero pressure resistivity at 250 K. The pressure drops by about 0.2 GPa between 300 K and 4 K.

At first, we discuss the zero pressure, ambient temperature case. The cross relaxation times, $T_x$ at zero pressure and 250 K in two κ-ET$_2$-Cl samples are 2.6±0.5 ns and 2.6±1 ns respectively. It is slightly longer, 2.9±0.5 ns, in a κ-ET$_2$-Br crystal. The systematic uncertainties of the fits are the main source of errors. From Eq. (2), we estimate the ambient temperature, zero pressure perpendicular resistivities:

$\rho_{\perp s}$=1.4 (1.5) Ωm in κ-ET$_2$-Cl (κ-ET$_2$-Br).

Here we used the hopping rates, $\nu_\perp$ =1.9×10$^8$ s$^{-1}$ (1.7×10$^8$ s$^{-1}$ ) for κ-ET$_2$-Cl (Br) and the density of states, $g(E_F)$ 4.4 states/eV per 2 ET (molecular dimer) from the calculations of [20]. The main uncertainty is in the value of the density of states, which is difficult to calculate and is ill defined in the bad metal state. There are only few absolute resistivity values reported in the literature. The 300 K $\rho_\perp$ values reported, 0.5 Ωm for κ-ET$_2$-Br [21] and 0.9 Ωm for κ-ET$_2$-Cl [22] are, however, consistent with $\rho_{\perp s}$.

In κ-ET$_2$-Cl, $T_x$ at ambient pressure increases with temperature between 300 K and 50 K similarly to the resistivity measured by [22]. (At lower temperatures spin-fluctuations complicate the ESR.) The similarity between the temperature dependence of $T_x$ and $\rho_\perp$ is evident [21] in κ-ET$_2$-Br, too. Like in the resistivity, there is a slight increase in the spin cross relaxation time between 250 K and 100 K that is followed by an abrupt decrease.

The parallel resistivity, $\rho_{//}$ of a κ-ET$_2$-Cl crystal from the same batch as the crystal used for the pressure dependent ESR was measured at 300 K as a function of pressure. The temperature dependence of $\rho_{//}$ was measured at pressures of 0.4, 0.64 and 0.96 GPa. The temperature and pressure dependence of the interlayer spin hopping time, $T_x$ follows roughly the parallel electric resistivity (Figs. 9, 10 and 11) normalized to the p=0 GPa, 250 K value.

The pressure dependence of the perpendicular hopping rate is similar to the perpendicular conductivity measured by Weiss [17], although $T_x^{-1}$ increases somewhat faster than $\sigma_\perp$. An increase of $\nu_\perp/\sigma_\perp$ with pressure is expected from eq.5.2 since the density of states normally decreases with pressure.

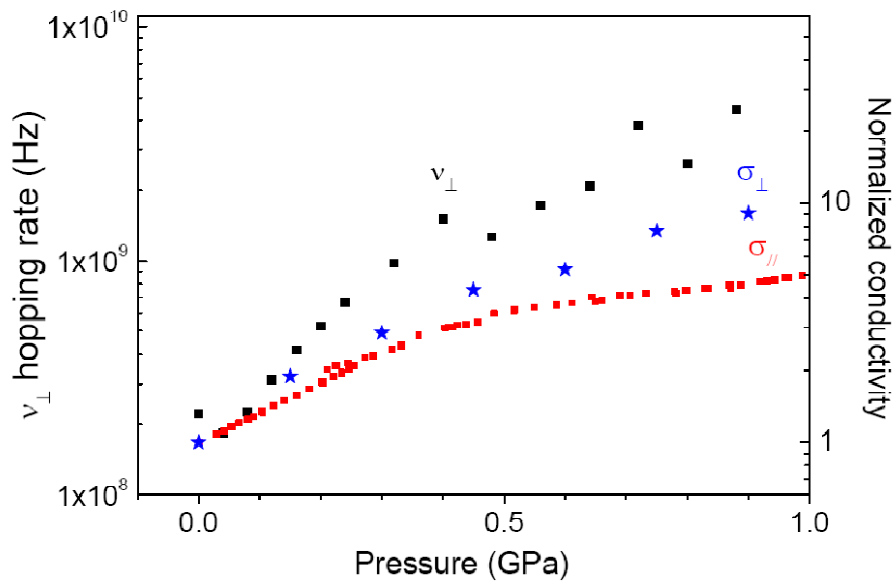

Fig.11. Comparison of the 250 K interlayer hopping rate $\nu_\perp$ (=1/(2$T_x$)) to the perpendicular conductivity, $\sigma_\perp$ (data from Ref. [17]) and parallel conductivity, $\sigma_{//}$ (present work) in κ-ET$_2$-Cl *versus* pressure. Conductivity data are normalized to zero pressure.

*5.3 Parallel conductivity in the bad metal phase*

The layered organic conductors, κ-ET$_2$-X are similar to the inorganic two dimensional systems in which the broadening and disappearance of long lifetime excitations at the Fermi surface with increasing temperature were directly observed by ARPES [23].

The measured high temperature resistivity anisotropy, $\rho_\perp/\rho_\parallel$ is typically $10^2$–$10^3$ in κ-ET$_2$-X above 50 K ([21], 1000), ([22],100). The corresponding parallel conductivity $\rho_\parallel$ of the order of $10^{-2}$ to $10^{-3}$ Ωm is extremely large. To see how large it is, we compare it to the resisitvity $\rho_{2D}=2\pi\hbar d/(e^2 k_F \lambda_c)$ of a 2D system with a free electron band structure. The Ioffe–Regel condition, i.e. a mean path equal to the molecular distance, $\lambda_c=a_0$ is the limiting condition for the metallic state of many normal metals [24]. Using this condition and a half filled band with the lattice parameters of κ-ET$_2$-X, we find $\rho_{2D} =1.8\times10^{-5}$ Ωm, a value more than two orders of magnitude smaller than the measured $10^{-2}$ to $10^{-3}$ Ωm. This implies that the Ioffe–Regel condition fails, and $\lambda_c \ll a_0$. Although it is notoriously difficult to measure the parallel conductivity in strongly anisotropic metals, an error of the order of $10^2$ or larger is unlikely and we conclude that the system is indeed a „bad metal". At 250 K the conductivity increases under pressure and tends towards a normal metal state. This increase is modest and the Ioffe–Regel condition is still not fulfilled at 1 GPa pressure.

The temperature dependence of the resistivity resembles the saturation phenomenon [24] observed in other metals. In these strongly interacting electronic systems, the phonon resistivity is no more linear with temperature and in some cases saturates at a high value with $\lambda_c \ll a_0$. In κ-ET$_2$-Cl at ambient pressure, electron-electron interactions are strong and the ground state is insulating. In the κ-ET$_2$-X metallic compounds the parallel resistivity at temperatures near $T_c$ is probably just compatible with the Ioffe–Regel condition. In κ-ET$_2$-Br a saturation resistivity is reached at 50 K, above this the conductivity is roughly temperature independent. At higher pressures saturation is not reached at 300 K and the resistivity has a nonlinear temperature dependence.

Merino and McKenzie have qualitatively well described the temperature dependence of the resistivity in κ-ET$_2$-X under pressure, using dynamical mean field theory in a strongly interacting half-filled band metal [25]. Assuming that pressure (or chemical pressure) changes the onsite Coulomb interaction, $U$, the resistivity saturates at high temperatures (like in κ-ET$_2$-Br) at a critical value of $U_c$. For weaker interactions the resistivity is metal-like (e.g. increases with temperature) while for $U>U_c$ it is insulator like. Using parameters relevant to κ-ET$_2$-Br, they find a saturation resistivity at $U=U_c$ of about $\rho_\parallel=10^{-4}$ Ωm.

*5.4 Blocking of the inter-layer hopping*

In strongly anisotropic conductors the interplane, „perpendicular" conductivity is often proportional to the in-plane, parallel conductivity in a wide temperature range. This is the case for κ-ET$_2$-Cl and Br between 50 and 300 K. The proportionality is natural in an anisotropic band conductor. However, at high temperatures the resistivity is very large and a Fermi liquid description of the perpendicular transport is certainly not valid. To explain a similar dilemma in the cuprate superconductors, Kumar [19] proposed that perpendicular hopping is blocked by rapid *intra-layer* scattering. Accordingly, an increase in parallel resistivity increases the perpendicular resistivity, so that the resistivity anisotropy remains unaffected.

At high temperatures, the intra-layer, parallel momentum scattering time, τ is much shorter than the characteristic time for tunnelling between layers, $\hbar/t_\perp$ where $t_\perp$ is the interlayer overlap integral:

$$\tau \ll \hbar / t_\perp . \qquad (3)$$

Frequent parallel scattering blocks tunnelling by a quantum mechanical effect; electron tunnelling from one molecular layer to the next restarts after every parallel scattering event. The tunnelling rate [19] is „blocked" by a factor of $2\tau/(\hbar/t_\perp)$:

$$\nu_\perp = 2\, t_\perp^2\, \tau / \hbar^2 . \qquad (4)$$

The perpendicular tunneling rate, $\hbar/t_\perp$ has been measured at low temperatures by angular magnetoresistance oscillations (AMRO) in some organic layered compounds. Inserting into Eq. (4) the typical value, $\hbar/t_\perp = 10^{-11}$ s [4], and $\nu_\perp = 1.9 \times 10^8$ s for κ-ET$_2$-Br at 250 K, the scattering time is $\tau = 1 \times 10^{-14}$ s.

According to Eqs.(4) and (2), the *perpendicular* conductivity is governed by the *parallel* scattering time, τ. However, the relation between the parallel conductivity, $\sigma_{//}$ and τ is not simple. It is tempting to follow the argument of Kumar [19], and equate τ to the scattering time $\tau_c$ of the conductivity, defined as

$$\sigma_{//} = (ne^2 \tau_c)/ m^* \qquad (5)$$

where n is the electron density and m* is the effective mass. Inserting $\tau = \tau_c$, the temperature independence of the conductivity anisotropy follows from Eqs. (2), (4) and (5):

$$\sigma_{//} / \sigma_\perp = \hbar^2/(2d^2\, m^* g(E_F)\, t_\perp^2 ). \qquad (6)$$

However, in the bad metal phase Eq. (5) does not hold. From the measured conductivity, Eq. (5) would imply $\tau_c \approx 10^{-16}$ s, i.e. $\hbar/\tau = 7$ eV, an unrealistic value, two orders of magnitude larger than the band width. Similarly, Eq. (6) is also inconsistent with the experiments. The same Eq. (6) is also valid for an anisotropic band conductor, it predicts a conductivity anisotropy of $10^5$–$10^6$, much larger than the observed $10^2$–$10^3$.

*5.5 Two-dimensional spin diffusion*

Spin diffusion is two-dimensional if spin memory of electrons is lost within a layer before hopping to an adjacent layer. This is clearly the case in κ-ET$_2$-Cl at ambient pressure and all temperatures. At room temperature $T_2$ is about 1 ns while $T_x$=2.6 ns. As a function of pressure, the 2D to 3D crossover at 250 K is at 0.2 GPa (Fig. 6). In κ-ET$_2$-Br the crossover to three-dimensional spin diffusion is at 40 K (Fig.8).

For possible applications, the 2D spin diffusion length, $\delta_{eff}$ in a single layer is important. In normal metals, $\delta_{eff} = [1/2(v_F^2 \tau_c T_1)]^{1/2}$ is determined by the intra-layer momentum scattering time and the spin–lattice relaxation time; in metals usually $T_1 \approx T_2$. In the normal metal phase, the Ioffe–Regel- condition holds at temperatures near $T_c$, $\tau_c > 10^{-14}$ s and $\delta_{eff} > 0.2$. We do not know what the corresponding expression is in the bad metal phase, where the quasiparticle picture fails. The low conductivity at high temperatures implies that the spin mean free path becomes smaller than 0.2 μm. On the other hand, the measured inter-layer hopping rate shows that the perpendicular conductivity is blocked by a slow process, $\tau=10^{-14}$ s, at high temperatures.

## 7. Conclusions

We have mapped the inter-layer spin hopping time and the intrinsic spin lifetime in the conducting phases of the temperature–pressure phase diagram of κ-ET$_2$-X. At high temperatures, above 50 K, spin diffusion is two-dimensional. The temperature and pressure dependence of the spin hopping time in κ-ET$_2$-X follows, at least qualitatively, the perpendicular resistivity $\rho_\perp$ as expected in a semiclassical picture where electronic and spin transport are tied together. The measured proportionality between $T_x$ and $\rho_\perp$ is well accounted for by typical density of states calculated in the literature. At high temperatures, the absolute value of $T_x$ and the proportionality of $\nu_\perp$ and $\rho_{//}$ are qualitatively reproduced by assuming the blocking mechanism suggested by Kumar and Jayannavar [19]. To understand the perpendicular hopping rate of $\nu_\perp=1.7\times10^8$ s for κ-ET$_2$-Br at 250 K by the blocking mechanism a parallel momentum scattering rate of $\tau=10^{-14}$ s is required. In the bad metal phase, however, the parallel conductivity is large and this scattering time cannot be the momentum relaxation time of coherent quasi particles as suggested in Ref. [19].

## Acknowledgements

The authors thank N. D. Kushch (Inst. of Problems of Chemical Physics, Chernogolovka, Russia) for instructions on crystal growth. We are thankful to Dr. Mátyás Czugler and Veronika Kudar (Chemical Research Inst. Budapest) for X-ray and K. Vad for secondary neutral mass spectrometry characterization of crystals. We are grateful for R. Gaál for resistivity measurements. This work was supported by the Hungarian National Research Fund OTKA PF63954, K68807 and the Swiss NSF and its NCCR "MaNEP". T. F. acknowledges financial support from the János Bolyai program of the Hungarian Academy of Sciences.